\documentclass[baaa]{baaa}

\usepackage[pdftex]{hyperref}
\usepackage{subfigure}
\usepackage{natbib}
\usepackage{helvet,soul}
\usepackage[font=small]{caption}

\begin{document}

%%%%%%%%%%%%%%%%%%%%%%%%%%%%%%%%%%%%%%%%%%%%%%%%%%%%%%%%%%%%%%%%%%%%%%%%%%%%%%
%                                                                            %
% Datos de la publicación, no deben ser cambiados.                           %
%                                                                            %
% Journal data, please do not change them.                                   %
%                                                                            %
%%%%%%%%%%%%%%%%%%%%%%%%%%%%%%%%%%%%%%%%%%%%%%%%%%%%%%%%%%%%%%%%%%%%%%%%%%%%%%

\journalvol{59}
\journalyear{2017}
\journaleditors{P. Benaglia, H. Muriel, R. Gamen \& M. Lares}

%%%%%%%%%%%%%%%%%%%%%%%%%%%%%%%%%%%%%%%%%%%%%%%%%%%%%%%%%%%%%%%%%%%%%%%%%%%%%%
%                                                                            %
%  Seleccione el idioma de su contribución: Recuerde que todos los           %
%  componentes del documento (titulo, texto, figuras, tablas, etc.)          %
%  deben estar en el mismo idioma.                                           %
%                                                                            %
%  Select the languague of your contribution: Please remember that all       %
%  document parts (title, text, figures, tables, etc.) must be in the        %
%  same languaje.                                                            %
%                                                                            %
%  0: Castellano / Spanish                                                   %
%  1: Inglés / English                                                       %
%                                                                            %
%%%%%%%%%%%%%%%%%%%%%%%%%%%%%%%%%%%%%%%%%%%%%%%%%%%%%%%%%%%%%%%%%%%%%%%%%%%%%%

\contriblanguage{1}

%%%%%%%%%%%%%%%%%%%%%%%%%%%%%%%%%%%%%%%%%%%%%%%%%%%%%%%%%%%%%%%%%%%%%%%%%%%%%%
%                                                                            %
%  Seleccione el tipo de contribución solicitada:                            %
%                                                                            %
%  Select the requested contribution type:                                   %
%                                                                            %
%  1: Presentación mural / Poster                                            %
%  2: Presentación oral / Oral contribution                                  %
%  3: Informe invitado / Invited report                                      %
%  4: Mesa redonda / Round table                                             %
%  5: Presentación Premio Varsavsky / Varsavsky Prize contribution           %
%  6: Presentación Premio Sahade / Sahade Prize contribution                 %
%  7: Presentación Premio Sérsic / Sérsic Prize contribution                 %
%                                                                            %
%%%%%%%%%%%%%%%%%%%%%%%%%%%%%%%%%%%%%%%%%%%%%%%%%%%%%%%%%%%%%%%%%%%%%%%%%%%%%%

\contribtype{1}

\thematicarea{7}

\title{EBL constraints with VERITAS gamma-ray observations}
%\subtitle{Instrucciones de estilo}

%%%%%%%%%%%%%%%%%%%%%%%%%%%%%%%%%%%%%%%%%%%%%%%%%%%%%%%%%%%%%%%%%%%%%%%%%%%%%%
%                                                                            %
%  Agregue un título corto para el encabezado de las páginas pares.          %
%                                                                            %
%  Add a short title to appear in the header of even pages.                  %
%                                                                            %
%%%%%%%%%%%%%%%%%%%%%%%%%%%%%%%%%%%%%%%%%%%%%%%%%%%%%%%%%%%%%%%%%%%%%%%%%%%%%%

\titlerunning{EBL constraints with VERITAS gamma-ray observations}

%%%%%%%%%%%%%%%%%%%%%%%%%%%%%%%%%%%%%%%%%%%%%%%%%%%%%%%%%%%%%%%%%%%%%%%%%%%%%%
%                                                                            %
%  Lista de autores. Los nombres de los autores deben estar separados por    %
%  comas, y deben tener el formato A.E. Autor (iniciales apellido(s);   sin  %
%  coma entre apellido e iniciales ni espacios entre las iniciales), y       %
%  con un simbolo "&" antes del último autor.                                %
%                                                                            %
%  Author list. Authors' names must be separated by commas, and stick to     %
%  the format A.E. Author (initials Family name -neither commas between      %
%  name and the initials nor blanks between the initials), and with an "&"   %
%  symbol before the last author.                                            %
%                                                                            %
%%%%%%%%%%%%%%%%%%%%%%%%%%%%%%%%%%%%%%%%%%%%%%%%%%%%%%%%%%%%%%%%%%%%%%%%%%%%%%

\author{Fernandez Alonso, M.\inst{1} for the VERITAS Collaboration}
\authorrunning{Fernandez Alonso et al.}

%%%%%%%%%%%%%%%%%%%%%%%%%%%%%%%%%%%%%%%%%%%%%%%%%%%%%%%%%%%%%%%%%%%%%%%%%%%%%%
%                                                                            %
% Por favor provea una dirección de e-mail de contacto para los lectores.    %
%                                                                            %
% Please provide a contact e-mail address for the readers.                   %
%                                                                            %
%%%%%%%%%%%%%%%%%%%%%%%%%%%%%%%%%%%%%%%%%%%%%%%%%%%%%%%%%%%%%%%%%%%%%%%%%%%%%%

\contact{mateofa@iafe.uba.ar}

\institute{Instituto de Astronomía y Física del Espacio (CONICET-UBA) 
}

%%%%%%%%%%%%%%%%%%%%%%%%%%%%%%%%%%%%%%%%%%%%%%%%%%%%%%%%%%%%%%%%%%%%%%%%%%%%%%
%                                                                            %
%  El resumen y el abstract son ambos obligatorios, independientemente del   %
%  lenguaje elegido.                                                         %
%                                                                            %
%  The Resumen and the abstract are both mandatory, regardless of the chosen %
%  language.                                                                 %
%                                                                            %
%%%%%%%%%%%%%%%%%%%%%%%%%%%%%%%%%%%%%%%%%%%%%%%%%%%%%%%%%%%%%%%%%%%%%%%%%%%%%%

\resumen{
La luz extragaláctica de fondo (EBL) contiene la radiación total emitida en procesos nucleares y de acreción desde la época de recombinación. La medición directa de este fondo es extremadamente difícil debido a la contaminación proveniente de la luz zodiacal. En cambio, la astronomía de rayos gamma ofrece la posibilidad de restringir de manera indirecta al EBL estudiando los efectos que la absorción de rayos gamma tiene en los espectros de fuentes detectadas en la banda de muy altas energías ($>$100 GeV). En general, los efectos de la absorción pueden apreciarse como un  {\it softening} en el espectro y/o cambios abruptos en el índice espectral de fuentes observadas. 
En este estudio se utilizan observaciones recientes de un grupo de blazares, realizadas con VERITAS, sobre las que se aplican dos métodos para obtener límites en las propiedades espectrales del EBL. Se presentan resultados preliminares que serán luego completados con observaciones mas recientes para mejorar las restricciones aplicadas el EBL. 
}

\abstract{
The extragalactic background light (EBL) contains all the radiation emitted by nuclear and accretion processes since the epoch of recombination. 
Direct measurements of the EBL in the near-IR to mid-IR waveband are extremely difficult due mainly to the zodiacal foreground light. Instead, gamma-ray astronomy offers the possibility to indirectly set limits on the EBL by studying the effects of gamma-ray absorption in the spectra of detected sources in the very high energy range (VHE: $>$100 GeV). These effects can be generally seen in the spectra of VHE blazars as a softening (steepening) of the spectrum and/or abrupt changes in the spectral index or {\it breaks}. 
In this work we use recent VERITAS data of a group of blazars and apply two methods to derive constraints for the EBL spectral properties. We present preliminary results that will be completed with new observations in the near future to enhance the limits on the EBL.
}

%%%%%%%%%%%%%%%%%%%%%%%%%%%%%%%%%%%%%%%%%%%%%%%%%%%%%%%%%%%%%%%%%%%%%%%%%%%%%%
%                                                                            %
%  Seleccione las palabras clave que describen su contribución. Las mismas   %
%  son obligatorias, y deben tomarse de la lista de la American Astronomical %
%  Society (AAS), que se encuentra en la página web indicada abajo.          %
%                                                                            %
%  Select the keywords that describe your contribution. They are mandatory,  %
%  and must be taken from the list of the American Astronomical Society      %
%  (AAS), which is available at the webpage quoted below.                    %
%                                                                            %
%  https://aas.org/authors/astronomical-subject-keywords-update-august-2013  %
%                                                                            %
%%%%%%%%%%%%%%%%%%%%%%%%%%%%%%%%%%%%%%%%%%%%%%%%%%%%%%%%%%%%%%%%%%%%%%%%%%%%%%


\maketitle

\section{Introduction}
\label{S_intro}

%\noindent A los participantes de la Reunión Anual No. 59 de la
%Asociación Argentina de Astronomía, llevada a cabo el pasado mes de septiembre.\\

The extragalactic background light (EBL) contains all radiation released from nuclear and accretion processes since the epoch of recombination. It consists essentially of all emitted and absorbed/re-emitted starlight accumulated over all redshifts. Understanding this background radiation is crucial to understand star formation processes and galaxy evolution models. So far, no direct detection has been achieved; the difficulty of this is due mainly to the zodiacal foreground light \citep{1998AAS...193.6202H}. However, throughout the last decade, upper and lower limits have been established using different methods, e.g. integrated galaxy counts from optical observations with the Hubble Space Telescope (\citealp{2000ApJ...542L..79G}; \citealp{2000MNRAS.312L...9M}), infrared observations using Spitzer Space Telescope (\citealp{2004ApJS..154...39F}; \citealp{2004ApJS..154...70P}) and the Infrared Space Observatory \citep{2002A&A...384..848E}. So far we have learned the EBL has a bimodal spectrum with one component peaking at $\sim1\mu m$ and another peaking at $\sim100\mu m$ \citep{2011ApJ...733...77O}. 

Gamma-ray astronomy offers the chance to indirectly set limits on the EBL by studying the effects of gamma-ray absorption in the spectra of detected sources in the very high energy range (VHE: $>$100 GeV). Gamma rays in the TeV regime have a high probability of interacting with background photons from the EBL via pair production \citep{1966PhRvL..16..252G}. The resulting leptons can interact via inverse Compton scattering with background photons as well, generating a cascade that results in the conversion of VHE photons into less energetic photons that can travel further \citep{1994ApJ...423L...5A}.
The cascade process results in an overall softening effect in the observed spectra of VHE energy sources. Moreover, the EBL spectral properties themselves can produce distinctive features in the observed spectra of VHE sources. In particular, recent EBL models, like \cite{2008A&A...487..837F}, predict an abrupt shift in the spectral index around 1 TeV that depends on the source distance, i.e. on the total EBL attenuation for a given source.

Blazars are a special type of active galactic nuclei (AGN) that have their jet pointing towards the Earth, and that present an unusually high TeV flux. The observed spectrum of these sources can be well characterized in the VHE range by a power law (e.g. \citealp{2001...UniverseinGamma}).
In this work we revisit a published method \citep{2011ApJ...733...77O} for analyzing blazar spectra and we use recent VERITAS \citep{2006APh....25..391H} data from the blazar 1ES1218+304 to test the method. This analysis will be part of a more complete EBL study that is being done by the VERITAS collaboration using more sources and methods.
                      
\section{Method}
Absorption of VHE photons can leave several traces in the observed spectra of blazars. In this work we explore two different methods to constrain the EBL using two different features present in observed blazar spectra: the softening of the spectral index (spectral shape method) and the spectral break around 1 TeV (spectral break method).
EBL constraints are derived by testing different models/realizations for the EBL spectral energy distribution (SED). For this study, a baseline model was assumed, following the shape outlined by lower limits gathered from the Hubble Space Telescope, the Spitzer Space Telescope and the Infrared Space Observatory. This baseline model is then used to generate different EBL scenarios following the {\it third order splines method} used in \cite{2007A&A...471..439M}. In this case the method is used to generate different SEDs by varying two wavelength values: $\lambda$=15$\mu$m and $\lambda$=1.6$\mu$m representing the mid and near IR regimes respectively.

%\begin{figure}[!t]
%  \centering
%  \includegraphics[width=0.45\textwidth]{Splines.png}
%  \caption{475 different EBL SED realizations generated with the third order splines method. Each curve is generated by varying the wavelenght values: $\lambda$=15$\mu m$ and $\lambda$=1.6$\mu m$.
%}
%  \label{splines}
%\end{figure}

\subsection{Spectral Shape Method} 

EBL absorption produces a steepening ({\it softening)} in the VHE part of the spectrum which depends both on the EBL properties and on the source distance. On the other hand, absorption has a minimal effect on the high energy (HE: 10 MeV$<$ E $<$100 GeV) part of the spectrum. Under the assumption that the intrinsic VHE spectrum of the source is a prolongation of the HE part, it is possible to test different EBL model-scenarios by using them to correct for EBL absorption in observed spectra and compare the resulting spectra to the corresponding HE spectra measured with Fermi-LAT. Using the EBL model it is possible to calculate the optical depth of a photon of a given energy and at a given redshift. The {\it intrinsic} spectrum is then calculated using the relationship

\begin{equation}
 \centering
\left( \frac{dN}{dE} \right)_{int}= \left( \frac{dN}{dE} \right)_{obs} e^{\tau\left( E,z\right) }
\label{eq_abs}
\end{equation}
where $\left( \frac{dN}{dE} \right)_{int}$ is the intrinsic spectrum, $\left( \frac{dN}{dE} \right)_{obs}$ is the observed spectrum, and $\tau\left( E,z\right) $ is the optical depth at energy E and source redshift z. $\tau$ is calculated for different EBL realizations generated with the splines method. The resulting intrinsic spectrum is then fitted by a power law and the model is accepted or rejected following the criterion
\begin{equation}
\centering
 |\Gamma_{TeV}-\Gamma_{GeV}|\leq \sqrt{\sigma^2_{GeV}+\sigma^2_{TeV}}
 \label{criterium}
\end{equation}
where $\Gamma_{TeV}$ and $\sigma^2_{TeV}$ are the calculated intrinsic spectral index and variance respectively, and $\Gamma_{GeV}$ and $\sigma^2_{GeV}$ are the Fermi spectral index and variance respectively.

\subsection{Spectral Break Method}

EBL absorption may also produce breaks in the observed spectrum. In particular, given the overall shape of the EBL, a break around 1 TeV is expected. The magnitude of these breaks increases with the source distance and depends on the EBL shape \citep{2008ICRC....3..981I}. Using a {\it test intrinsic spectrum}, the expected {\it observed} spectrum is calculated for different EBL realizations by using the inverse of equation \ref{eq_abs}. Gaussian fluctuations are added to the resulting points using a Normal distribution with a standard deviation equal to 25\% of each point's error bar. The resulting spectrum is then fitted with a broken power law with the form

\begin{equation}
  \frac{dN}{dE} = \begin{cases}
    N_{0}\left(\frac{E}{E_{break}} \right)^{-\Gamma_{1}}, & E \leq E_{break} \\
    N_{0}\left(\frac{E}{E_{break}} \right)^{-\Gamma_{2}}, & E > E_{break} 
  \end{cases} 
  \label{eq:broken}
\end{equation}
where $N_{0}$ is the normalization at the break energy $E_{break}$, $\Gamma_{1}$ and $\Gamma_{2}$ are the spectral indexes below and above the $E_{break}$ respectively and $E$ is the energy. For this particular study the break energy is fixed in 1 TeV and the fit functions are forced to match each other at this break point. The spectral break is then defined as

\begin{equation}
\Delta \Gamma = \Gamma_{1} - \Gamma_{2}
\label{delta_gamma}
\end{equation}

From here, the expected dependence of $\Delta \Gamma$ with redshift is estimated and then compared with the observed dependence. Doing this for different EBL realizations it is possible to test possible models by checking consistency between expected and observational results.

\section{Data Selection \& Analysis}

For the complete analysis 18 blazars were selected at various redshifts and with different spectral properties. All of them have been detected with a significance of more than 10 $\sigma$. The processing and reduction of the  TeV data is done with VERITAS own developed analysis software.
Fermi-LAT observations are used to obtain the source's spectra in the GeV regime.
At the moment the data is being processed and analyzed by members of the VERITAS collaboration using mainly the National Energy Research Scientific Computing Center (NERSC) cluster.

\section{Preliminary Results}

We have tested the de-absorption spectrum calculations and the methods on over 120 hours of 1ES1218+304 data to see the scripts are working properly. 

\begin{figure}[!t]
  \centering
  \includegraphics[width=0.5\textwidth]{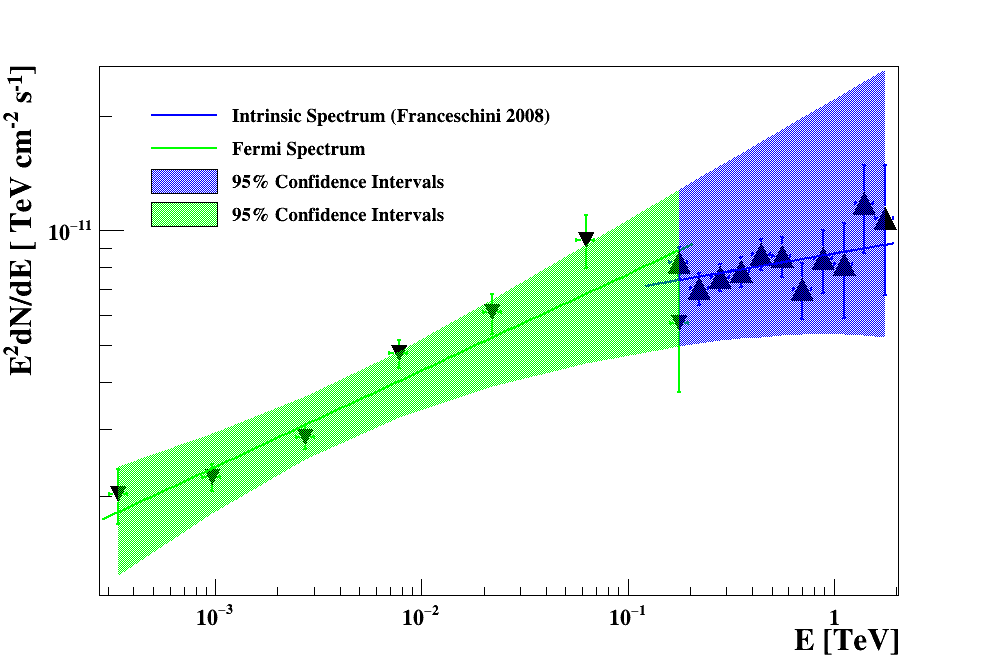}
  \caption{Combined GeV and intrinsic TeV spectrum calculated for 1ES1218+304. Each regime is fitted with a power law.}
  \label{1ES1218}
\end{figure}

Figure \ref{1ES1218} shows an example of the GeV and TeV spectra after de-absorption, each one fitted with a power law. The difference in the resulting spectral indexes $\Gamma_{GeV}$ and $\Gamma_{TeV}$ determines whether the model is accepted or rejected. Table \ref{tabla2} shows results obtained for three different EBL models as an example. 

\begin{table}[!t]
\centering

\begin{tabular}{ccc}
\hline\hline\noalign{\smallskip}
\!\! Model & \!\!\!\! N$\sigma$ & \!\!\!\! Condition  \!\!\!\!\\

\hline\noalign{\smallskip}
\!\! Franceschinni  &  0.65 & OK \\
\!\! Spline 1 & 0.24 & OK \\
\!\! Spline 2 & 3.33 & NO \\
\hline\noalign{\smallskip}
\end{tabular}
\caption{Differences between $\Gamma_{GeV}$ and $\Gamma_{TeV}$ in units of $\sigma$ for each EBL model. Models that fail the criterion given by equation \ref{criterium} are rejected.}
\label{tabla2}
\end{table}
 
\begin{figure}[!t]
  \centering
  \includegraphics[width=0.5\textwidth]{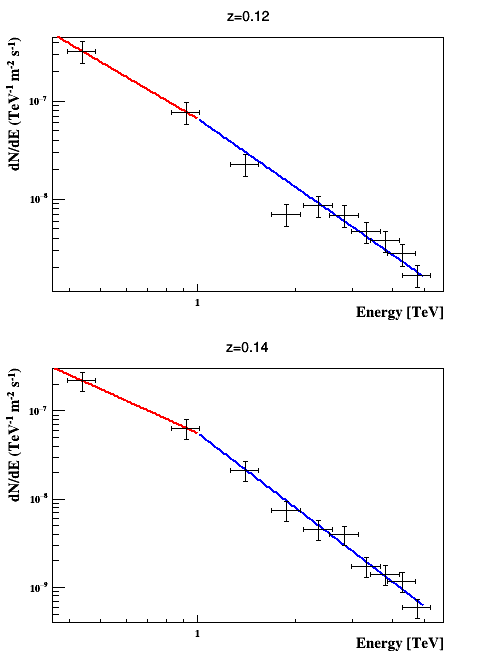}
  \caption{Expected TeV spectra for two different redshifts. Each spectrum is fitted with a broken power law below (red) and above (blue) the $E_{break}$=1 TeV, following equation \ref{eq:broken} }
  \label{break1}
\end{figure}

Figure \ref{break1} shows TeV spectra of two test sources at two different redshifts, fitted with a broken power law with a break at 1 TeV. It can be seen that the method

\begin{figure}[!t]
  \centering
  \includegraphics[width=0.5\textwidth]{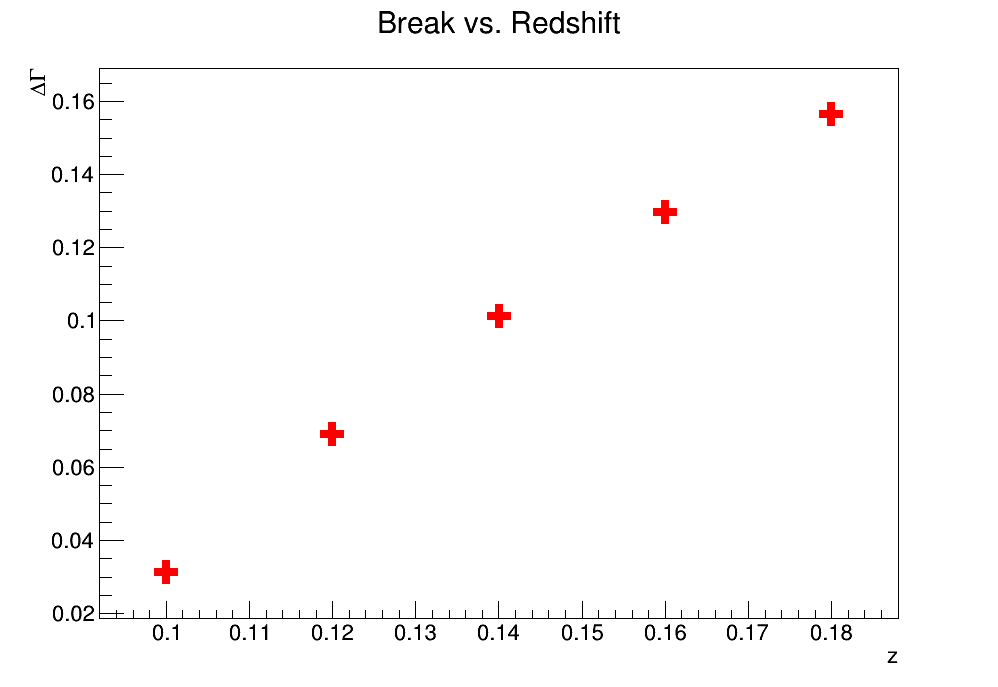}
  \caption{Expected trend of $\Delta \Gamma$ with z for a Crab-like spectrum.}
  \label{breakvsz}
\end{figure}

Figure \ref{breakvsz} shows an example of the expected trend of $\Delta \Gamma$ with redshift z obtained for a Crab-like spectrum\footnote{These results were obtained without considering any fluctuations, just to test the scripts were working properly.}. The behavior of $\Delta\Gamma$ with redshift will then be calculated for each considered source and compared to observations to constrain EBL models.

We tested and applied the {\it spectral shape} and {\it spectral break} methods on 1ES1218+304 data and checked scripts are working properly. These methods will be used on the new data once the processing and analysis phase concludes, and will hopefully derive more stringent constraints on the EBL spectral properties.

\begin{acknowledgement}
This research is supported by grants from the U.S. Department of Energy Office of Science, the U.S. National Science Foundation and the Smithsonian Institution, and by NSERC in Canada. We acknowledge the excellent work of the technical support staff at the Fred Lawrence Whipple Observatory and at the collaborating institutions in the construction and operation of the instrument.
\end{acknowledgement}

%%%%%%%%%%%%%%%%%%%%%%%%%%%%%%%%%%%%%%%%%%%%%%%%%%%%%%%%%%%%%%%%%%%%%%%%%%%%%%
%                                                                            %
%  Por favor no modifique las líneas de la bibliografía, salvo el nombre     %
%  el archivo de Bibtex con la lista de citas (sin la extensión .BIB)        %
%                                                                            %
%  Please do not modify the following lines, except the name of the Bibtex   %
%  file (whithout the .BIB extension)                                        %
%                                                                            %
%%%%%%%%%%%%%%%%%%%%%%%%%%%%%%%%%%%%%%%%%%%%%%%%%%%%%%%%%%%%%%%%%%%%%%%%%%%%%% 

\bibliographystyle{baaa}
\small
\bibliography{biblio_fernandez}
 
\end{document}